\documentclass[fleqn,usenatbib,useAMS]{mnras}

\usepackage{graphicx}   
\usepackage{amsmath}    
\usepackage{amssymb}    
\usepackage{multicol}        
\usepackage{bm}         
\usepackage{pdflscape}  

\newcommand{\hmsun}{h^{-1}{\rm ~M}_\odot}
\newcommand{\hmpc}{h^{-1}{\rm ~Mpc}}

\usepackage[T1]{fontenc}
\usepackage{ae,aecompl}

\usepackage{txfonts}

\title[Correlations of extended structures]{Spatial correlations of extended cosmological structures}

\author[Santucho et al.]{\parbox[t]{\textwidth}{%
Santucho, V. \thanks{Contact e-mail:
   \href{mailto:santucho@oac.unc.edu.ar}{santucho@oac.unc.edu.ar}}, 
Luparello, H. E., Lares, M., Lambas, D. G., Ruiz, A. N. \&  Sgr\'o, M. A.  }\vspace{0.2cm}\\ 
Instituto de Astronom\'{\i}a Te\'{o}rica y Experimental, CONICET-UNC, 
and Observatorio Astron\'{o}mico de C\'{o}rdoba, 
Argentina}

\date{Released June 2019 }
\pagerange{\pageref{firstpage}--\pageref{lastpage}} \pubyear{2019}

\def\LaTeX{L\kern-.36em\raise.3ex\hbox{a}\kern-.15em
    T\kern-.1667em\lower.7ex\hbox{E}\kern-.125emX}

\begin{document} 
\label{firstpage} 
\maketitle 

\begin{abstract} 
Studies of large--scale structures in the Universe, such as superstructures or cosmic voids, have been widely used to characterize the properties of the cosmic web through statistical analyses. 
On the other hand, the 2--point correlation function of large--scale tracers such as galaxies or halos provides a reliable statistical measure. However, this function applies to the spatial distribution of point--like objects, and therefore it is not appropriate for extended large structures which strongly depart from spherical symmetry. 
Here we present an analysis based on the standard correlation function formalism that can be applied to extended objects exhibiting arbitrary shapes. Following this approach, we compute the probability excess $\Xi$ of having spheres sharing parts of cosmic structures with respect to a realization corresponding to a distribution of the same structures in random positions. 
For this aim, we identify  superstructures defined as Future Virialized Structures (FVSs) in semi-anaytic galaxies on the MPDL2 MultiDark simulation.  We have also identified cosmic voids to provide a joint study of their relative distribution with respect to the superstructures.
Our analysis suggests that $\Xi$ provides useful characterizations of the large scale distribution, as suggested from an analysis of sub--sets of the simulation. 
Even when superstructure properties may exhibit negligible variations across the sub--sets, $\Xi$ has the sensitivity to statistically distinguish sub--boxes that departs from the mean at larger scales. 
Thus, our methods can be applied in analysis of future surveys to provide characterizations of large--scale structure suitable to distinguish different theoretical scenarios.
\end{abstract} 

\begin{keywords}
   large scale structure of Universe --
   cosmology: observations -- 
   methods: statistical -- data analysis
\end{keywords}

\section{Motivation} \label{S_intro}

A network of interconnected filaments, walls, and dense clusters
surrounded by large subdense void regions provides a useful
description of the large--scale structure of the Universe
\citep[e.g.][]{Einasto:1997, Colless:2001, Jaaniste:2004,
Einasto:2006, Abazajian:2009, Einasto:2014}.
Galaxy surveys of large extension revealed this complex pattern
\citep{York:2000,Colless:2001}, which is also  supported by
cosmological N-body simulations evolving a CDM--type primordial power
spectrum \citep{Frisch:1995,Bond:1996,Sheth:2003,Shandarin:2004}.
In this scenario, the largest mass concentrations coincide with the
intersections of walls and filaments conforming the locations of
clusters and superclusters of galaxies.
These massive objects have been studied and identified with a large
variety of methods, such as linking Abell cluster positions
\citep{Zucca:1993,Einasto:1997}, or directly from wide--area surveys
of the large-scale galaxy distribution
\citep{Shectman:1996,Colless:2001,Stoughton:2002}.
More recently, supercluster catalogs were constructed by combining new
methodologies of clustering analysis
\citep{Einasto:1997,Einasto:2006,Liivamagi:2012,Nadathur:2014,Chow:2014}
providing a deeper insight into the nature, extension and distribution
of the largest bound structures of the Universe.
In the $\Lambda$CDM scenario, the present and future dynamics of the
Universe is dominated by an accelerated expansion, hence an
appropriate definition of large--scale structures can be obtained by
the requirement of present overdense regions that will be bound and
virialized systems in the future
\citep{Busha:2005,Dunner:2006,Araya:2009,Luparello:2011}. 
Studies of the evolution of the superstructure network show that
although some small systems may have  undergone merger events with
other superstructures during their past assembly history, this is not
a dominant factor on the global evolution.
Rather, the main transformation processes takes place inside the
systems \citep{Einasto:2019}.
Different attempts to provide useful characterization
of the large--scale structure have been performed either using the
galaxy distribution \citep{Tegmark:2004, Scottez:2016} as well as 
studying large scale galaxy system
properties such as superclusters, clusters, filaments and voids 
\citep{Colberg:2000,Brodwin:2007,Chan:2014}.
Early studies included counts in cells moments \citep{White:1979},
topological genus of smoothed density distributions \citep{Gott:1986},
Minkowsky functionals \citep{Schmalzing:1996}, structure functions on
minimal spanning trees \citep{Colberg:2007}, wavelet methods, etc. 
More recently, other approaches to extract high order information have
focused on non-linear transforms such as a logarithmic transformation
method \citep{Wang:2011}.
It is worth to consider that large--scale structure studies, mainly
through detailed analysis of the two--point correlation function, have
given an important support in the settlement of the standard
cosmological concordance model. 
High order statistics has been used in present
survey analyses \citep {Verde:2002, Gaztanaga:2005, Baugh:2004, Croton:2007} since higher
order correlations can give useful information about 
the complex filamentary structure.

Although straightforward calculation of high order correlations is
possible and has been computed in observations and in numerical
simulations \citep[see for instance,][and references
therein]{Slepian:2015}, beyond the three point function estimations
become challenging not only from the computational point of view, but
also offers model interpretation in terms of covariances for large
number of data bins.
However, it should be recalled the fact that this statistic considers
a distribution of point sources, not extended objects with arbitrary
shapes where the center definition cannot be properly addressed.
When such structure sizes and mean separations are comparable, the
2--point correlation function formalism cannot be used
straightforward.
In cases with spherical symmetry, as in the case of voids, this is
overcame by comparing to a random distribution of points that take
into account void sizes.
It is worth noticing that although properties of voids have been used
to test different cosmological models as well as modified gravity,
over-dense superstructures such as the Future Virialized Structures
\citep[FVSs,][]{Luparello:2011} have been less studied, mainly because
of their complex spatial shape. 
\citet{Lares_whales:2017} studied the spatial correlations between voids and
superstructures by defining a first modified version of the 
traditional 2--point correlation function.
They quantified the cross correlation of voids relative to FVSs estimating the 
probability of finding a randomly placed sphere containing simultaneously void and FVS.
That allowed them to measure the tendency of voids to be located at a given distance 
from a superstructure. In this framework, considering the random spheres as a 
suitable tool, we propose an alternative method, based on the formalism of 
the 2--point correlation function, that can be applied to measure clustering 
on extended and non--symmetrical systems. Also, as the process includes the comparison 
with random data sets the results are easier to assess. Moreover, this allows 
us to avoid systematic and intrinsic bias associated with any particular 
sample of structures.\\ 
This paper is organized as follows: in the next section, we describe the data and methods applied to the identification of FVSs and voids.
Then, in Section \ref{sec_method} we introduce the definition of the structure correlation function, along with some tests intended to qualitatively interpret its behaviour.
In Section \ref{S_results} we compute the structure auto--correlation for FVSs, also exploring the effects of cosmic variance. Besides, we analyse the FVS-Void structure cross--correlation.
Finally, in Section \ref{S_conclusions} we present a summary and discussion of our results.

\section{Data}  \label{S_data}

The present work is based on the public semi--analytic catalogues
MultiDark--Galaxies \citep{Knebe:2018}, available at the CosmoSim
(\url{https://www.cosmosim.org}) and Skies \& Universes
(\url{https://www.skiesanduniverses.org}) databases. Particulary, we
use the catalogue MDPL2-SAG\footnote{doi:10.17876/cosmosim/mdpl2/007},
which was constructed populating the MDPL2 dark matter haloes with the
semi--analytic model of galaxy formation and evolution \citep[SAG,
][]{Cora:2018}.
The MDPL2 simulation follows the evolution of 3840$^{3}$ particles in
a cubical volume of side-length 1 h$^{-1}$ Gpc, and form part of the
MultiDark suite of simulations \citep{Riebe:2013, Klypin:2016}.
The adopted cosmology consists of a flat $\Lambda$CDM model with the
\textit{Planck} cosmological parameters: $\Omega_{m}$ = 0.307,
$\Omega_{b}$ = 0.048, $\Omega_{\Lambda}$ = 0.693, $\sigma_{8}$ =
0.823, n$_{s}$ = 0.96 and a dimensionless Hubble parameter h = 0.678
\citep{Planck_cosmo:2015}.
The mass resolution is m$_{p}$ = 1.51 $\times$ 10$^9\hmsun$ per dark
matter particle.
Haloes and subhaloes have been identified with \textsc{rockstar} halo
finder \citep{Behroozi:2013a} and \textsc{ConsistentTrees} has been
used to construct the merger trees \citep{Behroozi:2013b}.
Here we aim at presenting a useful statistic for large--scale
structures, that hopefully could be applied to current or future
observational galaxy catalogues.
We have then selected a sample of semi--analytical galaxies brighter
than $-20.6$ magnitudes in the r--band, so that the numerical density
of the sample is roughly that of volume limited samples of galaxies.

\subsection{FVSs catalogue}\label{FVS_data}

As stated above, recent superclusters catalogues were constructed
mainly based on the smoothed density field approach
\citep[e.g.][]{Einasto:2007,CostaDuarte:2011,Liivamagi:2012,Einasto:2014}.
However, these systems are still in the beginning of the virialization
process, thus the density threshold applied to delineate the
superclusters has a certain degree of freedom.
Taking this into account, \citet{Luparello:2011} used a criterion
motivated by dynamical considerations to calibrate this value and
presented a catalogue of so called Future Virialized Structures.
The procedure relies in combining the luminosity density field method
\citep{Einasto:2007} with the theoretical criterion for the mass
density threshold of \citep{Dunner:2006}.
The method starts with a smooth luminosity density field obtained by
convolving the spatial distribution of the galaxies with a kernel
function weighted by galaxy luminosity. 
The effect of the smoothing depends on the shape and size of the
kernel function, an Epanechnikov kernel of radius \mbox{r$_{0} = $ 8
$\hmpc$}.
The resulting density field is then mapped into a grid with a
resolution given by cubes of 1 $\hmpc$ side.
Then, a percolation method allows to select the highest luminosity
density groups of cells.
These isolated over-densities are the basis of the large structures
that will become virialized systems in the future.
To this end, the luminosity--density threshold adopted \mbox{(D$_{T}$
= $\rho_{lum} \slash \bar{\rho}_{lum}$ = 5.5)} is calibrated, assuming
a constant mass to luminosity relation and using the minimum mass
overdensity necessary for a structure to remain bound in the future
\citep{Dunner:2006}. 
A lower limit for the total luminosity of a structure is also applied
at L$_{structure}$ $>$ 10$^{12}$L$_{\odot}$, avoiding contamination
from smaller systems.
These criteria provides a suitable compromise of high completeness and
low contamination.
The described procedure was applied to the MDPL2-SAG catalogue, where
we identify 3219 FVSs, with volumes between 10$^{2}$ and $\simeq$
10$^{4}(\hmpc)^{3}$ and  luminosities ranging from \mbox{10$^{12}$ to
$\simeq$ 10$^{14}$ (L$_{\odot}$ h$^{-2}$)}.

\subsection{Void catalogue} \label{VOIDS_data}

We follow the void identification algorithm as described in
\cite{ruiz_clues_2015}, which is a modified version of the algorithms
presented in \cite{padilla_spatial_2005} and
\cite{ceccarelli_voids_2006}.
The method starts with an estimation of the density field performed
with a Voronoi tessellation over the denstity tracers (SAG galaxies
in this case).
Firstly, under--dense regions are obtained by selecting all Voronoi
cells below a threshold density. These cells are then used as center
candidates for under--dense larger regions.
The integrated density contrast in spherical regions around these
cells, $\Delta(r)$, is then computed at increasing values of radius
$r$. Then, void candidates are selected as the largest of each set of
overlapping spheres that satisfy the condition $\Delta(R_{\rm
void})<-0.9$.
Since this procedure can lead to spherical voids with edges that do
not precisely fit with the surrounding structures, centers are
displaced randomly so that the spheres are allowed to grow. The
procedure of re--centering void locations provide systems whose
borders are in better agreement with the local density field
surroundings.
We end up with a void catalog comprising all the largest under--dense,
non overlapping spheres, with void radii $R_{\rm void}$ equals to
those of re--centered spheres.

Physical insight on the nature and evolution of void properties can be
derived from voids surroundings since the hierarchy of voids arises by
the mass assembly in the growing nearby structure network
\citep{sheth_hierarchy_2004, paranjape_more_2012}.
In this scenario, some voids remain as underdense regions, while other
voids evolve by collapsing onto themselves together with the
surrounding dense structures.
Both possible behaviors, either void expansion or void collapse, are
determined by the global surrounding density:
large underdense regions with surrounding overdense shells are likely
to be squeezed as the surrounding structures behave as a {\it
void-in-cloud} system, voids in an environment more similar to the
mean background density will expand and remain as underdense regions
following the so called {\it void-in-void} mode. 
Cumulative radial density profiles have proved to be a useful proxy to
predict these two evolutionary modes
\citep{ceccarelli_clues_2013,paz_clues_2013,ruiz_clues_2015} and can
be used to perform a classification of voids: those with a smoothly
rising but still under--dense integrated radial profile resembling
void-in-void systems, dubbed R-type voids; and those embedded in
globally over--dense regions, consistent with the void--in--cloud
mode, defined as S-type (standing for shell--surrounded) voids.
The final catalogue comprises 17975 voids for the full MDPL2-SAG box,
with radii in the range 7--30 $\hmpc$.

\section{Method} \label{sec_method}

\begin{figure}
\centering \includegraphics[width=0.45\textwidth]{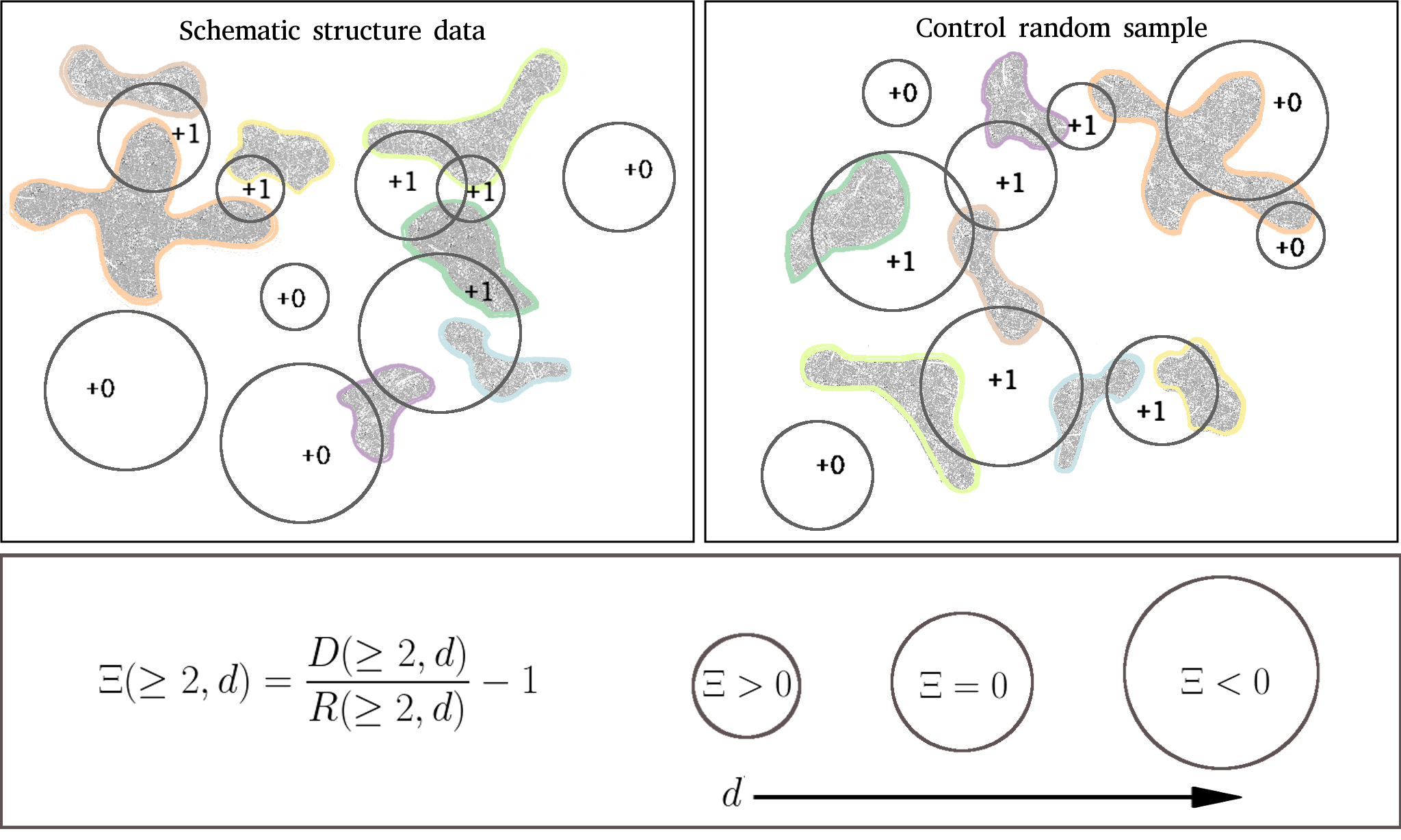} 
   \caption{Simplified schematic representation of the procedure. 
   Upper panels: shaded regions display a simulated distribution of non-symmetrical
   extended objects (left) and their corresponding randomized sample, both in position and orientation (right). 
   The same structure in both panels is represented by the same  contour colour.
   Some random spheres are displayed overlapping the structures and the numbers inside of the
   spheres indicate the contribution to $D(\geq 2, d)$ and $R(\geq 2, d)$.
   Lower panel: The three spheres on the right,  
   qualitatively shows the signs of $\Xi(\geq2,d)$ computed by Eq. \ref{eq_auto}, also shown in the left lower panel. These spheres correspond to the scales outlined in the top panels.
} 
   
\label{F1} \end{figure}

Here, we describe a novel tool to characterize the clustering of
non--symmetrical large--scale systems.
We base our method on the 2--point correlation function formalism,
modified for application to non--symmetrical, extended systems.
As a first step, and following the idea of quantifying the different
spatial distribution of superstructures from a randomly placed sample
of the same systems, we use a reference randomly distributed sample by
shuffling the actual position and orientation of the structures, 
with the restriction of no superposition. 
Thus, we end up with random catalogues that have the 
same number of structures and the same volume than the 
data.
We repeat this procedure four times with different
random seeds in order to generate different realizations of the random catalogues. 
We find that the results for the four random realizations are statistically indistinguishable.
We apply this procedure for the generation of random catalogues of both cosmic voids and overdense structures in the following sections.
In order to quantify the different clustering of the actual
distribution and that derived from randomly shuffling the
structure positions, we use the fraction of randomly placed spheres of
a given diameter (a proxy for the scale of the correlation) that
overlap with the structures in both data and randomly placed
structures.
If there is a clustering signal in the distribution of
superstructures, then these fractions would differ in such a way that
can be quantified as a function of the scale with statistical
significant estimators.
The probability for a sphere of diameter $d$ to be in contact with at
least $k$ structures, $P(\ge k,d)$, can be approximated by the
fraction of random spheres of that diameter that are superimposed to
$k$ or more than $k$ structures:
\begin{equation} P( \ge k, d) = N( \ge k, d) / N(d), \end{equation}
where $N(d)$ is the total number of random spheres.
This quantity may differ for superstructures in the dataset (in our
case, the simulation) and in the set with randomly placed
superstructures.
Then, we define a structure correlation function $\Xi(\ge k, d)$ of
order $k$ as the excess probability of finding a sphere of diameter
$d$ partially or totally overlapping at least $k$
different structures w.r.t. an equivalent arrangement with random structures:
\begin{equation} 
P_D(\ge k, d)  = P_R(\ge k ,d) \,(1+\Xi(\geq k,d)).
\end{equation}
Within this scheme, we define: $D(\geq k , d)$ as the number of random
spheres that contains parts of at least $k$
different structures, in the simulation box, and $R(\geq k, d)$ as the number of
random spheres that contains parts of at least $k$ different
structures, in the random sample. 
We notice that when a structure contacts a random sphere in more than one location 
this intersection is considered a single overlap.\\
Also, we fix the total number of spheres 
for each diameter bin N(\textit{d}) to be the same in the simulation box and in the random sample. 
With these quantities, we estimate the "structure auto--correlation
function" of order 2 ($k \ge 2$) as:
\begin{equation} \label{eq_auto} 
\Xi(\geq2,d) = \frac{D(\geq 2, d)}{R(\geq 2, d)} -1, 
\end{equation}
and the "structure cross correlation function" as:
\begin{equation} \label{eq_cross}
\Xi_{VS}(\geq2,d) = \frac{D_{VS}(\geq 2, d)}{R_{VS}(\geq 2, d)} -1, 
\end{equation}
where subscripts $V$ and $S$ represent a cosmic void and an overdense
structure, respectively.
Here we consider spheres in contact with at least one void and at least one FVS.\\ 
Fig. \ref{F1} shows a scheme of our procedure, where the top
panels sketch a simulated distribution of non-symmetrical extended objects (left) and its corresponding randomized sample, both in position and orientation (right). 
Contours of different colours are used to easily recognize the randomized structures.
Three different sphere diameters are shown in the scheme and
the numbers ($+0$ and $+1$) superposed to each of them
indicate the contribution to the counts $D(\geq 2, d)$ or $R(\geq 2, d)$. It can be seen that spheres in contact with two or more structures contribute with one count and spheres that overlap with less than two 
structures do not contribute to the counts.
The lower panel shows qualitatively the three different scales from the upper panels in which the auto--correlation give positive, zero and negative values respectively. 
The three scales were arbitrarily chosen to illustrate possible behaviors of the correlation function computed by Eq. \ref{eq_auto}.

\subsection{Tests with simple geometrical models}
In order to address the performance of the $\Xi(\geq2,d)$ correlation
we have constructed five different models of structures in the same
grid of the simulation data.
The models follow
simple distribution prescriptions and geometrical shapes, so that the behavior
of the correlation function can be assessed.
We acknowledge the restriction of $30 \hmpc$ as the minimum distance between the structure centers.
This constraint is applied to all models in order to avoid overlapping of structures.
We adopted two structure shapes, one is a set of cells consistent with a sphere of radius $8 \hmpc$ (adding up 2106 $\hmpc^3$),
and the other, a prism with 26, 9, 9 $\hmpc$ a side.
The volumes of these structures are approximately the mean volume of the
FVSs identified in the MDPL2-SAG simulation.
One of the structure models consists of spheres (S-model) and the remaining models of prisms (P-models).
The total volume and number of structures (set as 8673 objects) is the same for all the five models.

\begin{figure}
\centering \includegraphics[width=0.45\textwidth]{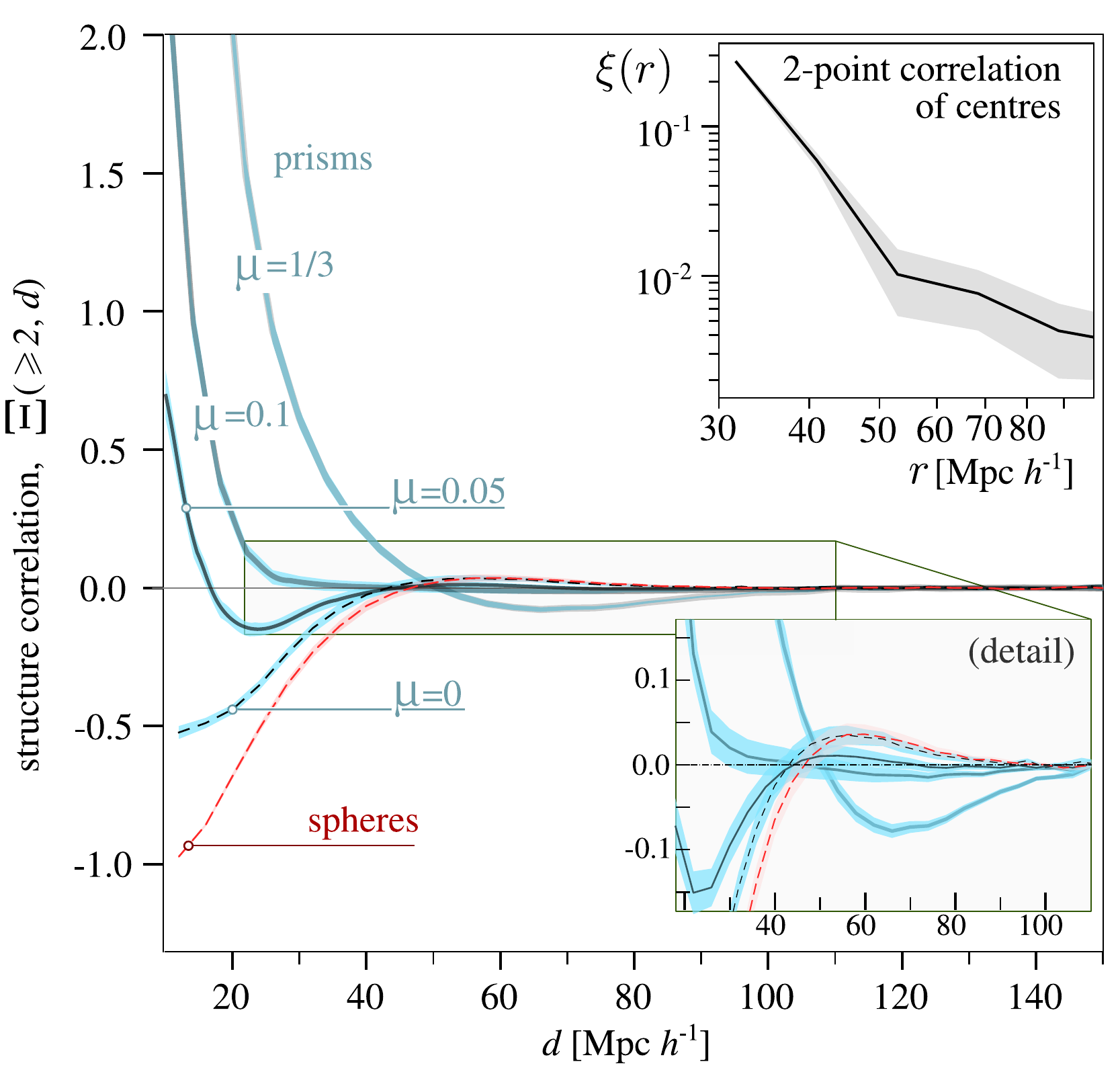}      
   \caption{Structure auto--correlation function
   in toy models. The red dashed line corresponds to S-model
   auto--correlation $\Xi_{S}(\geq2,d)$. Solid and dashed cyan lines show the
   prisms auto--correlation $\Xi_{P}(\geq2,d)$ for different $\mu$
   values, labeled in the Figure.
   For a better comparison, the lower inset displays a zoomed region 
   of the structure auto--correlation functions.     
   The upper inset shows the 2--point correlation function of the centers used to place the structures in the S-model and P-model with $\mu$=0. Shaded regions correspond to Jackknife uncertainties.
   } 
\label{F1b} \end{figure}

The spheres in the S-model are distributed so that they show a
clustering signal in the 2--point correlation function of their
centers, as shown in the upper inset panel of Figure \ref{F1b}.
In the construction of the P-models, we randomly chose a fraction $\mu$ of the
S-model centers to place a pair of nearby and parallel prisms, and a fraction of
1-2$\mu$ to place one isolated prism.
The prisms are oriented along a randomly chosen axis.
We have adopted P-models with $\mu$ values of 0, 0.05, 0.1, and 1/3, 
and where
the pairs of prisms have a constant separation of $2 \hmpc$.
We notice that in the cases where $\mu \ne 0$ the prism centers have a very different exclusion restriction.
We acknowledge that the original exclusion restriction is modified since pairs of prisms are arranged so that their separation is significantly lesser than the $30 \hmpc$ of the original prescription. 
These new distribution of prisms also affects the original two-point correlation function.
We stress the fact that in all cases we impose the condition of no overlapping of structures, consistent with the definition of FVSs.
The results are shown in Figure \ref{F1b}. The effect of FVS shapes can be seen by comparing the S--model (red dashed line) and the P--model with $\mu$=0 (blue dashed line), where the largest difference occurs at small scales due to the effects of the elongated prism shapes.
The larger extension of the prism--shaped structures in one of the axis with respect to spherical-shape structures, make more probable that a random sphere contact two or more different prisms.
For the P--model with $\mu$=0, the negative correlation values of $\Xi_{P}(\geq2,d)$ correspond to the imposed exclusion distance between centers. At increasing separations beyond this exclusion scale, the effects of the original standard two point correlation of the centers can also be distinguished in $\Xi_{P}(\geq2,d)$ as positive values at large scales.
Beyond  $\gtrapprox 100 \hmpc$ both the standard and the $\Xi(\geq2,d)$ correlations tend to zero.\\
It can also be appreciated the effects of the introduction of a significant clumping at small 
scales in the different P--models corresponding to $\mu$ values of 0.05, 0.1, and 1/3.  
At small separations, the large positive values of $\Xi_{P}(\geq2,d)$ correspond to the excess of close pairs of prisms.
We notice that even a small fraction of prism pairs, $\mu$=0.05, gives a positive correlation at small scales, showing negative values at intermediate separations due to the imposed exclusion condition. 
We notice that at larger scales, the positive correlation of the original distribution of centers provides positive values of $\Xi_{P}(\geq2,d)$ as well.
For higher values of $\mu$ (i. e. 0.1, 1/3), this positive signal is absent due to the fact that there are practically no traces of the original spatial distribution.
%

\section{Analysis and results} \label{S_results}

\begin{figure*}
\centering \includegraphics[width=\textwidth]{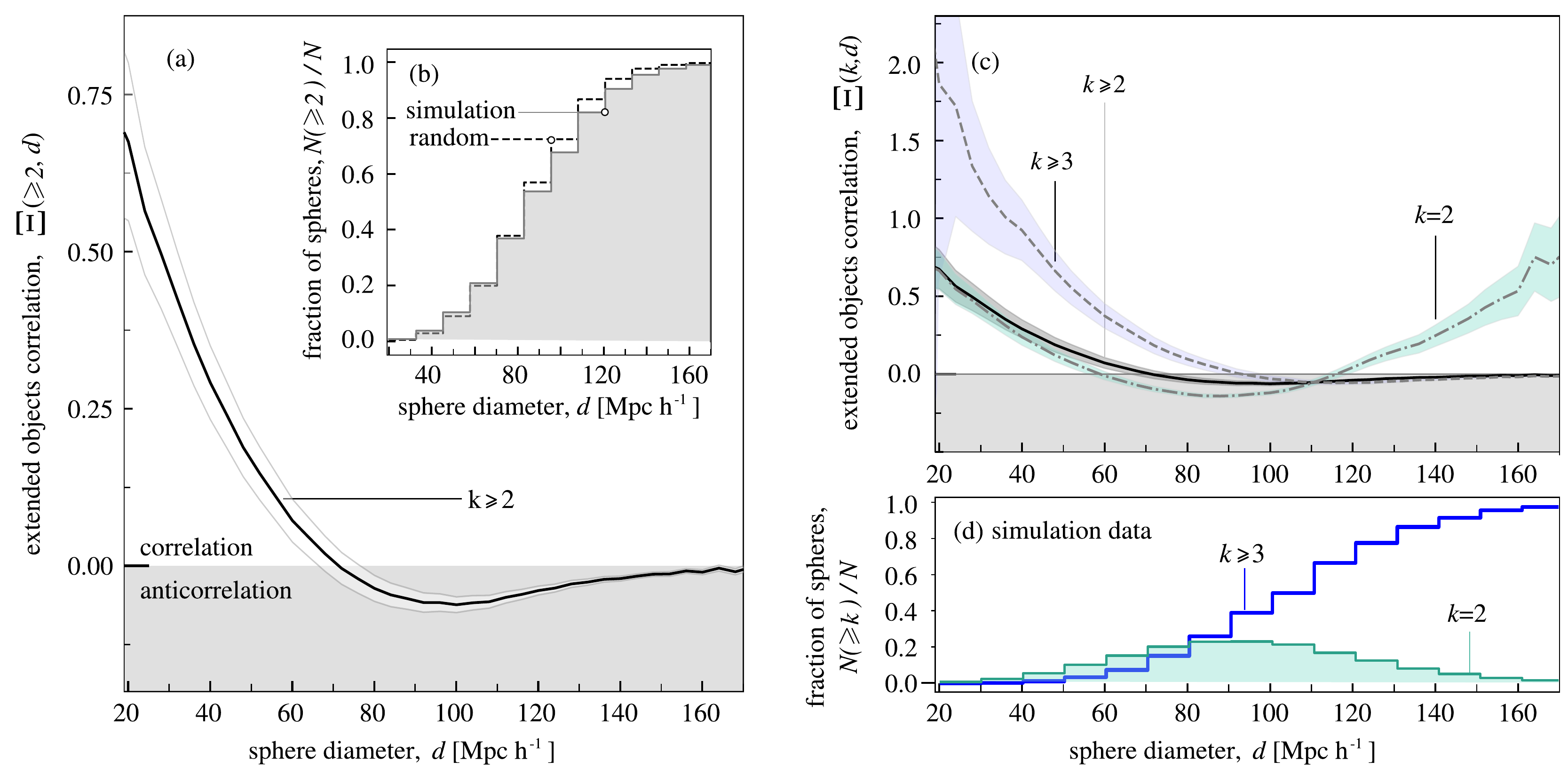}
\caption{FVS Structure auto--correlation function $\Xi_{FVS}(\geq2,d)$
   in MDPL2-SAG catalogue. Panel (a): $\Xi_{FVS}(\geq2,d)$ for
   MDPL2-SAG. The inset (panel b) shows the distribution of the number of
   spheres $D(\geq 2, d)$ (solid line) and $R(\geq 2, d)$ (dashed
   line) involved in the computation of $\Xi_{FVS}(\geq2,d)$, as
   exemplified in Eq. \ref{F1}. Panel (c): Contribution of the
   $\Xi_{FVS}(2,d)$ (thin line) and the $\Xi_{FVS}(\geq3,d)$
   structure correlation function (thick line) to the total
   structure auto correlation (lightgray line). In both panels, shaded
   regions correspond to Jackknife uncertainties. Panel (d) shows the
   number of spheres involved in the determination of the
   $\Xi_{FVS}(2,d)$ and $\Xi_{FVS}(\geq3,d)$
   structure correlations in diameter bins for the simulation box.} 
\label{F2} \end{figure*}

Using Eq. \ref{eq_auto}, we compute the FVS structure
auto--correlation function ($\Xi_{FVS}(\geq2,d)$) in the MDPL2-SAG
catalogue and show it in panel (a) of Fig.~\ref{F2}.
To that end, we generate random spheres and estimate the correlation
from the fractions of spheres that contain structures.
For smaller scales, where the variance is large due to the structure,
we use a larger number of spheres.
The choice of the distribution of the spheres radii was made so that 
their total volume is equivalent to 100 times the simulation box volume.
As it can be seen in the plots, due to the intrinsic dispersion of $\Xi_{FVS}(\geq2,d)$ at small scales
the Jackknife uncertainty estimates are significantly higher than at larger scales.
Histograms in the inset (panel b) represent the fraction of the number of
spheres that are in contact with at least two structures relative to
the total number of spheres, as a function of sphere diameter. The
solid lines correspond to the simulation box, and the dashed lines to
the control sample.
As it can be seen (panel a), $\Xi_{FVS}(\geq2,d)$ exhibits a positive
correlation at scales lesser than 70 $\hmpc$. 
Beyond this scale, FVS spatial distribution shows an anti--correlation
up to scales of roughly 150 $\hmpc$, which smoothly approaches to zero
at large scales, consistent with a random distribution.
The strong positive correlation signal at small separations can be
understood in terms of the arrangement of FVSs within the large--scale
structure network as, for instance, large--scale filaments.

According to our definition, any configuration of two or more
structures in contact with a randomly placed sphere 
has the same contribution to our correlation measure,
regardless of the common volume fraction or of the 
number of contributing structures.
In order to deepen our understanding of the FVS structure
auto--correlation function, we have considered separately the number
of structures contained in the random spheres to compute
$\Xi_{FVS}(k,d)$.
The panel (c) of Fig.~\ref{F2} shows the FVS
auto--correlation function considering $k=2$ and $k\ge3$ separately. 
The gray solid line represents the
correlation of FVSs computed in Sec.~\ref{sec_method},
and the thick solid line measures the contribution of spheres 
containing only two different FVSs $(\,\Xi_{FVS}(2, d)\,)$.
Similarly, the thin solid line represents $\Xi_{FVS}(\ge 3, d)$, 
corresponding to the contribution of three or more FVSs.
We acknowledge that the $\Xi_{FVS}(2, d)$ and $\Xi_{FVS}(\ge 3, d)$
terms dominate the signal of the full structure correlation at
different scales.
Although the mean number of structures contained in the
spheres scales with diameter, the contribution of pairs in
sufficiently large spheres will be negligible since it is increasingly
difficult to find regions containing only two structures.
As it can be seen in panel (c) of Fig.~\ref{F2}, at scales 
beyond $\simeq$ 100 $\hmpc$ the the main
contribution is due to the $\Xi_{FVS}(\ge 3,d)$ term.
To further understand these results, we show in panel (d) Fig.~\ref{F2}
the histograms corresponding to the fraction of spheres (with respect
to the total number of spheres with the same diameter range)
containing $k=2$ (thin solid line) and $k \ge 3$ (thick solid line) structures where the $k \ge 3$ contribution clearly dominates beyond $100 \hmpc$.

\subsection{Subsamples and cosmic variance}
\label{SS_variance}

\begin{figure}
\includegraphics[width=0.45\textwidth]{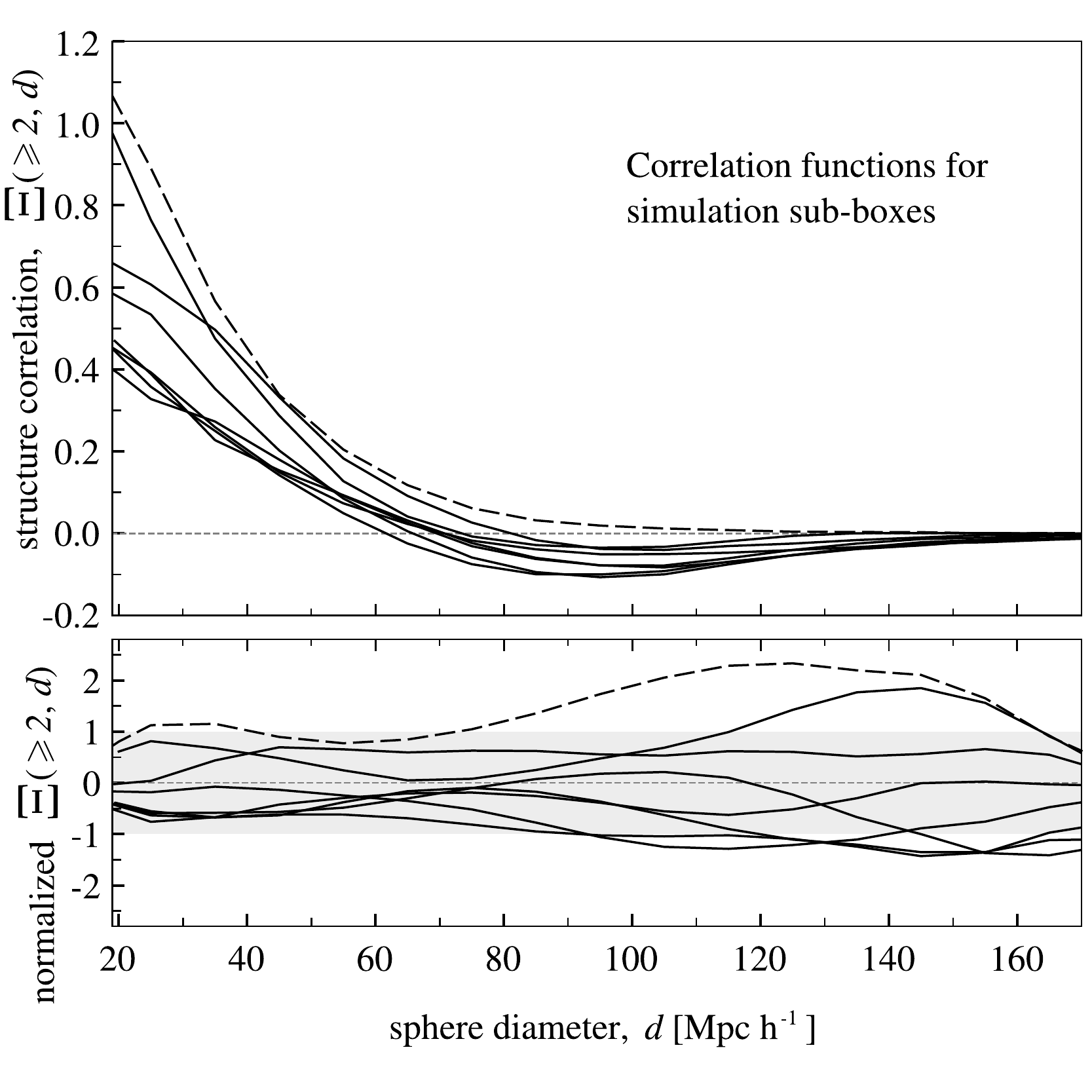}      

\caption{FVS Structure auto correlation function for MDPL2-SAG sub--boxes. 
The dashed line corresponds to the box with the particular behavior mentioned in section \ref{SS_variance}. 
The lower panel shows the difference of the structure correlation of each sub--box with respect to the full box in units of the mean correlation variance of each sub--box. The shaded region represents 1 standard deviation from the full box correlation.
The box highlighted in the upper panel is shown with dashed line.}
\label{F3} \end{figure}

\begin{figure}
\includegraphics[width=0.45\textwidth]{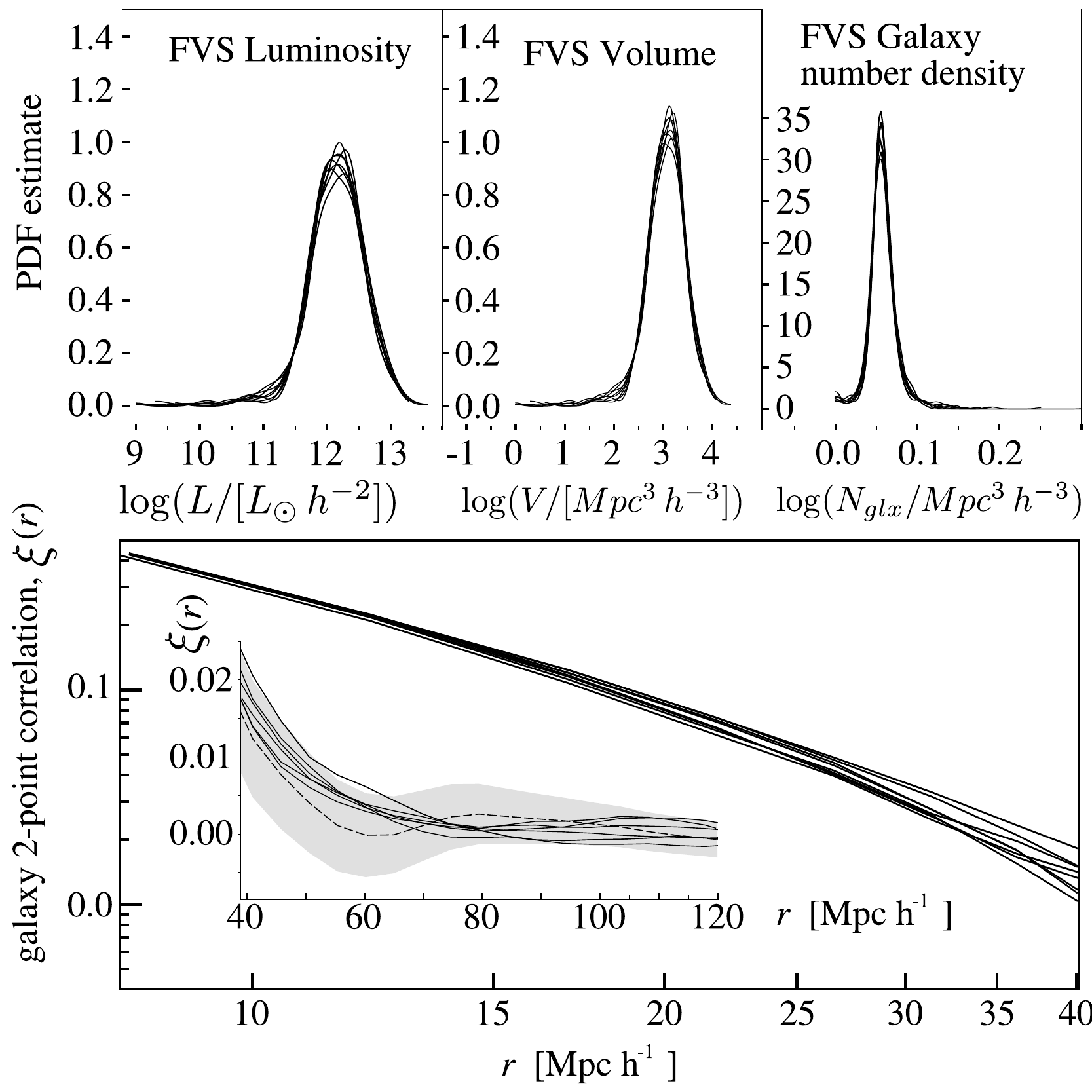}
\caption{Probability density of FVSs luminosity (left panel), volume
   (middle panel) and galaxy number density (right panel)
   distributions for  FVSs in MDPL2-SAG sub--boxes. Lower panel:
   2--point correlation function of galaxies for MDPL2-SAG sub--boxes. 
   The inset shows 2--point correlation function beyond 40 $\hmpc$ in linear scales.
   The dashed line corresponds to the mentioned sub--box in section \ref{SS_variance} and the gray region shows its Jackknife uncertainty.}
\label{F4} \end{figure}

With the aim at exploring the effects of cosmic variance on the
structure correlations we define 8 non overlapping sub--boxes of 500
$\hmpc$ side of the MDPL2-SAG catalogue. 
For each sub--box we compute the FVS structure auto correlation
function. The results obtained are shown in the upper panel of Fig.~\ref{F3}.
The lower panel shows the difference of the structure correlation of each sub--box with respect to the full box in units of the mean correlation variance of each sub--box. The gray region corresponds to 1 standard deviation from the full box correlation.
As it can be noticed, there is one distinct sub--box in which the
structure correlation function shows a highest correlation
with respect to the other sub--boxes although we notice that its significance is only at the 2 sigma level. 
To analyse this feature in further detail, and in particular its
possible dependence on differences in characteristics of FVSs across
the sub--boxes, we have analysed the distribution function of several
intrinsic properties of FVSs in the MDPL2-SAG sub--boxes.
The upper panel of Fig.~\ref{F4} presents the probability density of
FVS luminosities (left panel), FVS volumes (middle panel) and FVS multiplicity, i.e., the number of 
SAG galaxies within the limiting magnitude (right panel) in the MDPL2-SAG sub--boxes.
These FVS properties are remarkably similar for the different sub--boxes,
indicating that the observed clustering differences can not be endowed
to differences in FVS intrinsic characteristics.
We have also studied the luminosity distribution of the galaxies
within each sub--box and the results are in complete agreement among
the sub--boxes. 
Besides these analysis, we have also analysed whether differences in
the structure correlation function could be directly related to
variance in the galaxy auto correlation function. 
Hence, we have analysed the standard 2--point galaxy--galaxy
correlation function in the different sub--boxes.
The lower panel in Fig.~\ref{F4} shows the standard 2--point correlation
function of galaxies for the 8 MDPL2-SAG sub--boxes.
It can be seen that all sub-boxes show very similar correlation functions at scale $<$ 40 $\hmpc$ although we notice that beyond this scale the same box exhibiting the higher FVS structure correlation also departs from the trend of the remaining sub--boxes. This can be seen in the inset of the lower panel of the Fig.~\ref{F4}, where the above--mentioned sub-box is represented with dashed line.
However we stress the fact that, in both FVS Structure and standard correlation functions, all determinations are within cosmic variance uncertainties.
We have also checked if the superstructure correlation excess of the sub--box 
with highest correlation is also present in both $\Xi_{FVS}(2,d)$ and
$\Xi_{FVS}(\geq 3,d)$ finding a consistent excess in both cases. 
Thus, to conclude from this analysis, we find that the larger excess of
superstructure correlation of this sub--box relies on very
large-- scale correlations rather than on intrinsic FVS properties.
%

To further explore the possible origin of the structure correlation
excess in the sub--box with highest correlation, we imposed a lower threshold density for FVS identification (see Sec.~\ref{S_data}) and studied the resulting FVS properties in each MDPL2-SAG sub--box.
This analysis shows that, particularly in this sub--box, the
currently identified FVSs were part of a larger superstructure.
We argue that this fact can explain the actual clustering excess with
respect to the other boxes.
Thus, this test shows that the FVS structure auto--correlation
function contains information on the properties of the underlying
density field and large--scale structure.

\subsection{FVS--Void structure cross--correlation}

\begin{figure*}  
\includegraphics[width=\textwidth]{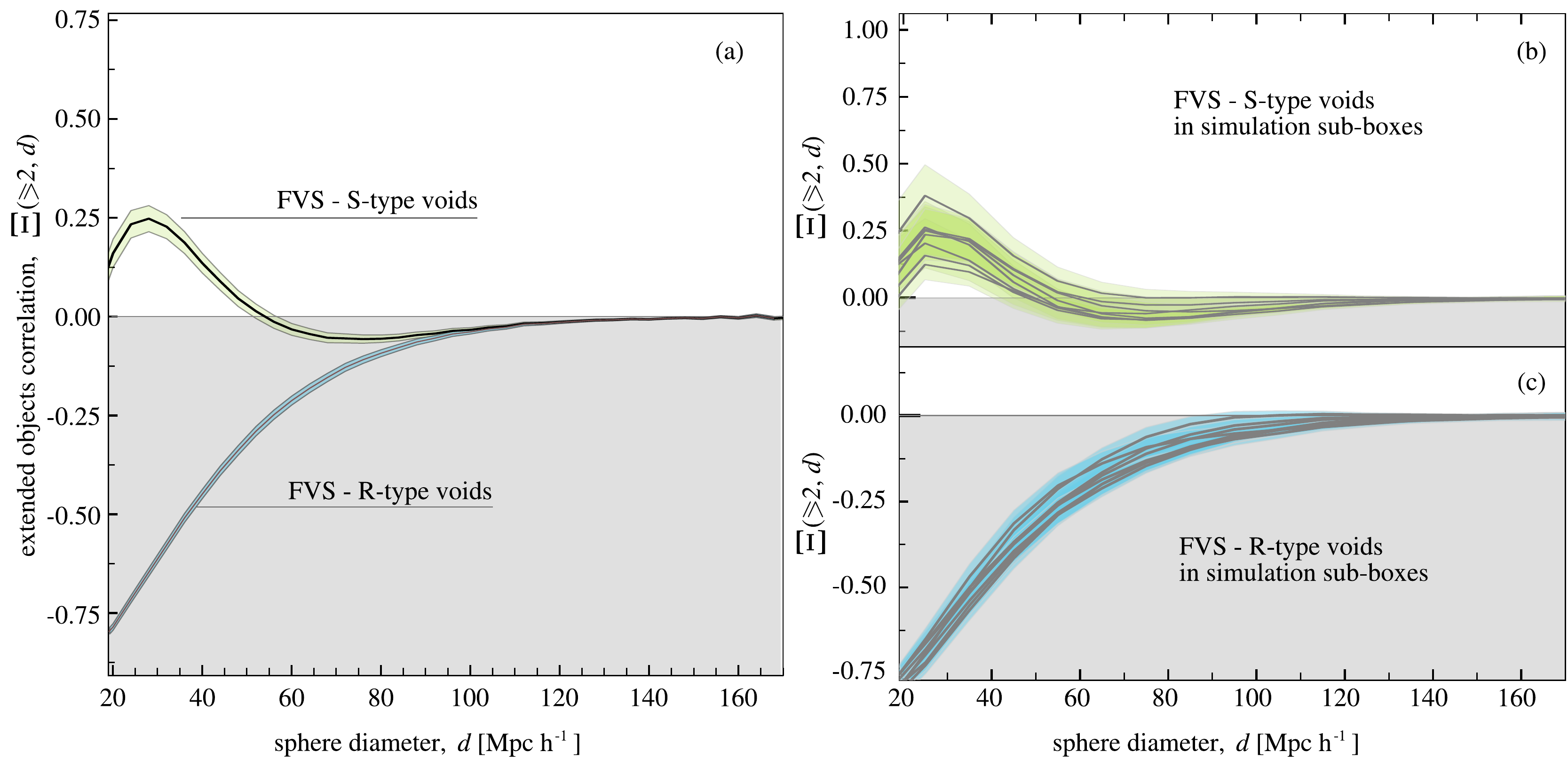}
\caption{FVS--VOID Structure cross--correlation for the total
MDPL2-SAG catalogue. Left panel shows FVS--Void
structure cross--correlations for FVSs and S-- and R--type void samples.
Right panels (b and c) show the same correlation as in panel (a)
but for the 8 simulation sub--boxes in order to visualize the cosmic variance.Shaded regions in both panels correspond to Jackknife
uncertainties within each sub--box.}
\label{F5} \end{figure*}

In previous sections we have studied the correlated positions of
superstructures by means of an appropriate statistics.
Since the
cosmic web comprises under--dense regions as well as superstructures,
it is useful to analyse their joint distribution.

In this subsection we compute the structure cross--correlation
function between FVSs and Voids, $\Xi_{SV}(\geq2,d)$, by applying the
Eq. \ref{eq_cross} for FVSs (S) and Voids (V).
We consider the spheres
that are in contact with at least one void and at least one FVS.
The left panel of Fig.~\ref{F5} shows the FVS--Void structure
cross--correlation function for the total MDPL2-SAG catalogue.
The upper line corresponds to $\Xi_{SV}(\geq2,d)$ between
FVSs and S--type voids, and the bottom line to
$\Xi_{SV}(\geq2,d)$ between FVSs and R--type voids.
We can distinguish two opposite behaviours: S--type voids are
correlated with FVSs while R--Type voids tend to avoid FVSs proximity,
consistently with previous studies \citep{Lares_whales:2017}.
As stated in \citet{Lares_whales:2017}, this is somewhat related to the
selection of voids according to their environment.
Moreover they found that this behavior has also dynamical 
implications when considered in the context of the large--scale 
structure, manifested on the infall of voids which are close 
to FVSs independently of void type.
Given the results obtained for the cosmic variance of the structure auto correlation
function of FVSs  $\Xi_{FVS}(\geq2,d)$ from Sec.~\ref{SS_variance}  we have also
explored $\Xi_{SV}(\geq2,d)$.
The results are shown in the right panel of Fig.~\ref{F5} where it can be seen that the sub--box with the highest auto correlation also shows a slightly excess of the structure cross--correlation amplitude between FVSs and S--type Voids. 
However, for the all sub-boxes the structure cross--correlations are within cosmic variance uncertainties.
Thus, we envisage that cross--correlations between superstructures and
voids may also provide useful statistical characterizations of the
spatial distribution in next generation of large--scale surveys.

\section{Conclusions} \label{S_conclusions} 

In this paper we have addressed the computation of spatial
correlations of large--scale cosmic structures with arbitrary shape.
These structures are determinant in the distribution of galaxies at
large scales and being highly asymmetrical, standard N--point
correlation function methods are not suitable for statistical studies.
Our method is based on the probability excess $\Xi(k,d)$ of finding
spheres of diameter $d$ overlapping with cosmic structures of a given
species with respect to a randomized distribution of the same
structures. 
We have applied this procedure to superstructures and voids identified
in the MDPL2-SAG catalogue and we have tested its performance in
deriving useful characterizations of the large scale distribution
properties. 
We have analysed cosmic variance effects in sub--sets of the MDPL2-SAG
catalogue on $\Xi(\geq2,d)$ and on the standard galaxy correlation
function statistics. 
We find that both $\Xi(\geq2,d)$ and the standard correlation function
are suitable to detect departures from the correlation of the full box
due to cosmic variance.
Thus, the structure correlation can provide a useful complementary
characterization of the large--scale mass distribution in 
future large galaxy surveys.
We stress the fact that, even when the superstructure properties do
not strongly vary among the different sub--boxes, the 
structure correlation function and the standard correlation function
of halos are still sensitive to the cosmic variance.\\
Regarding to FVS--VOID structure cross--correlation, 
our results are consistent with those presented in
\citet{Lares_whales:2017} where FVSs and R-type (S-type) voids have
negative (positive) correlations. This provides further support to the
ability of our methods to produce useful characterizations of
the large-scale structure.
We argue that, since FVSs emerge as features of the cosmic
web such as knots in the filamentary structure of the supercluster--void
network, their formation and evolution are intertwined, thus allowing
to use superstructures as an alternative proxy for the statistical
description of the large--scale network of the distribution of
galaxies.

\section*{acknowledgements}
This work was partially supported by the Consejo Nacional de
Investigaciones Cient\'{\i}ficas y T\'ecnicas (CONICET), and the
Secretar\'{\i}a de Ciencia y Tecnolog\'{\i}a, Universidad Nacional de
C\'ordoba, Argentina.
Plots were made using Python software and postprocessed with
Inkscape.
This research has made use of NASA's Astrophysics Data System. 
The CosmoSim database used in this paper is a service by the
Leibniz-Institute for Astrophysics Potsdam (AIP). The authors
gratefully acknowledge the Gauss Centre for Supercomputing e.V.
(\url{https://www.gauss-centre.eu}) and the Partnership for Advanced
Supercomputing in Europe (PRACE, \url{https://www.prace-ri.eu}) for
funding the MultiDark simulation project by providing computing time
on the GCS Supercomputer SuperMUC at Leibniz Supercomputing Centre
(LRZ, \url{https://www.lrz.de}).

\bibliographystyle{mnras}
\bibliography{main}
\bsp  
\label{lastpage}

\end{document}